# Non-zero Integral Spin of Acoustic Vortices and Spin-orbit Interaction in Longitudinal Acoustics

Wei Wang*, Yang Tan*, Jingjing Liu[†], Bin Liang[†], Jianchun Cheng[†]

Key Laboratory of Modern Acoustics, MOE, Institute of Acoustics, Department of Physics, Collaborative Innovation Center of Advanced Microstructures, Nanjing University, Nanjing 210093, People's Republic of China

*These two authors contributed equally to this work.

[†]Correspondence and requests for materials should be addressed to Jingjing Liu (email: liujingjing@nju.edu.cn), Bin Liang (email: liangbin@nju.edu.cn), or Jianchun Cheng (email: jccheng@nju.edu.cn)

## Abstract

Spin and orbital angular momenta (AM) are of fundamental interest in wave physics. Acoustic wave, as a typical longitudinal wave, has been well studied in terms of orbital AM, but still considered unable to carry non-zero integral spin AM or spin-orbital interaction in homogeneous media due to its spin-0 nature. Here we give the first self-consistent analytical calculations of spin, orbital and total AM of guided vortices under different boundary conditions, revealing that vortex field can carry non-zero integral spin AM. We also introduce for acoustic waves the canonical-Minkowski and kinetic-Abraham AM, which has aroused long-lasting debate in optics, and prove that only the former is conserved with the corresponding symmetries. Furthermore, we present the theoretical and experimental observation of the spin-orbit interaction of vortices in longitudinal acoustics, which is thought beyond attainable in longitudinal waves in the absence of spin degree of freedom. Our work provides a solid platform for future studies of the spin and orbital AM of guided acoustic waves and may open up a new dimension for acoustic vortex-based applications such as underwater communications and object manipulations.

# Main

Spin and orbital angular momenta (AM) are of fundamental interest in both classical and quantum wave physics[1-5]. Orbital AM, originating from the circulation of the phase gradient and manifesting as a helical wavefront, can be carried by both transverse and longitudinal waves[6]. Spin AM, on the other hand, is produced by the circular or elliptical polarization of waves and characterized by the local rotation of vector fields, limiting its in-depth studies only in transverse wave systems currently. In optics, orbital and spin AM have been extensively studied in the past few decades both in theory and application, and have become ubiquitous in many areas of modern optics[7-10]. However, acoustic waves propagating in fluids turn out to be longitudinal waves. Hence, instead of spin AM, considerable efforts have been dedicated to the study of acoustic orbital AM in the form of vortex beams both theoretically and experimentally[11-16], which have shown great potential in a wide range of acoustics-based application scenarios[17-21].

Recently, it is found that nontrivial polarization and spin properties can also exist in longitudinal acoustic waves that can also be treated as longitudinal vector waves corresponding to spin-0 longitudinal phonons in terms of vector velocity field[2,22-26]. However, the spatial integration of spin AM density of the acoustic vortex beam is believed always equal to zero[2], directly resulting in the lack of spin degrees of freedom. Consequently, spin-orbit interaction (SOI), an intriguing phenomenon with orbital and spin AM conversion that is observed in various ways and has aroused enormous interests in optics[27,28], is considered to be beyond attainable in longitudinal acoustics previously[29].

In this article, we present the first analytical calculations of the spin and orbital AM for the cylindrical guided modes using self-consistent quantum-like representations. The schematic of eigenmodes in the acoustic cylindrical waveguide is shown in Fig. 1a. Different from the fact that the spatially integrated spin AM values equal zero in free space complying with its spin-0 nature[2], we find that the integral spin AM of cylindrical guided modes can be non-zero and proportional to the kinetic-Abraham total AM density at the boundary, which is not found in optics or other systems. Our results reveal the fundamental features of spin and orbital AM, universal for acoustic modes in various kinds of boundary conditions. In addition, Minkowski and Abraham momentum are two forms of momentum of light in a medium[30,31], whose debate has lasted a century-long time, and in free space these two variables are equivalent when considering integral over the whole space[32]. Here, we also analyze the acoustic kinetic-Abraham and canonical-Minkowski AM under different boundary conditions, and prove that only the former is conserved with the corresponding symmetries. The conserved total AM, as well as the spin degree of freedom provided by non-zero spin AM carried by cylindrical guided modes, make the observation of SOI of vortices in longitudinal acoustic systems no longer out of reach. Hence, based on this, we give the first experimental demonstration of SOI of longitudinal acoustics using a cylindrical waveguide with varying cross-sections.

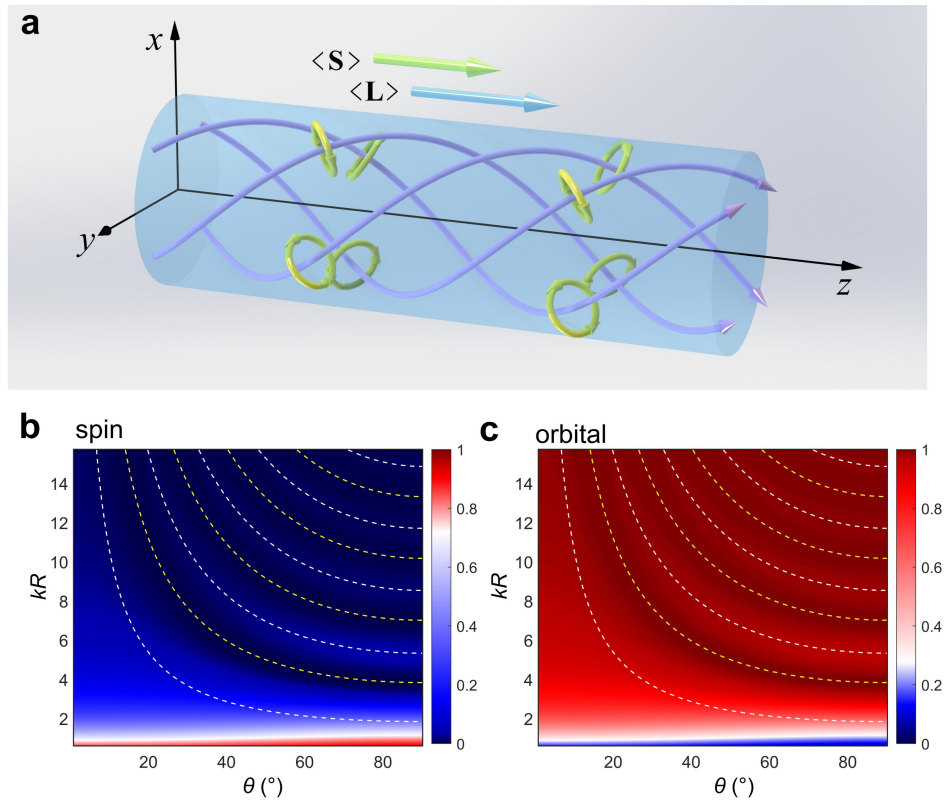

**Fig. 1| Schematic of spin and orbital AM of cylindrical guided mode and normalized spin and orbital AM distribution with different boundary conditions.**

**a**, Schematic of spin and orbital AM of cylindrical guided mode, where yellow and purple arrows in the waveguide represent the origin of the spin and orbital AM, respectively; while green and blue arrows out of the waveguide represent the integral spin and orbital AM, respectively. Normalized **b**, spin and **c**, orbital AM distribution varying with the paraxial degree and radius of cylindrical waveguide with $\theta = \sin^{-1}(k_r/k)$. White and yellow dashed lines correspond to absolutely soft and absolutely hard boundary conditions, respectively.

## Dynamical properties of acoustic waves

The main dynamical properties of acoustic waves are energy, momentum, and AM. The cycle-averaged energy density and energy flux density in a monochromatic acoustic field are expressed as

$$W = \frac{1}{4}\left(\beta|p|^2 + \rho|\mathbf{v}|^2\right), \quad \mathbf{I} = \frac{1}{2}\mathrm{Re}\left(p^*\mathbf{v}\right), \tag{1}$$

where $\rho$ and $\beta$ are the mass density and the compressibility, respectively, while $p$ and $\mathbf{v}$ are the acoustic pressure and velocity field, respectively.

Acoustic waves can be described in quantum-like terms by the four component "wave function" $\psi = (p, \mathbf{v})^T$, and the energy density can be regarded as the local expectation value of the energy operator $\omega$, $W = (\psi|\omega|\psi)$, where the inner product $(\psi|\psi)$ is defined with the scaling coefficients $\beta/4\omega$ and $\rho/4\omega$ at the pressure and velocity degrees of freedom, respectively[2]. Here we use the same presentation as that in electromagnetism $\mathbf{S} = (\psi|\hat{\mathbf{S}}|\psi)$, where $\hat{\mathbf{S}} = \begin{bmatrix} 0 & 0 \\ 0 & \hat{\mathbf{s}} \end{bmatrix}$ with $\hat{\mathbf{s}}$ being spin-1 operator. Canonical Minkowski-type momentum, spin, orbital, and total AM densities can be obtained by corresponding operators in such presentation as[33]

$$\mathbf{P} = \frac{1}{4\omega}\mathrm{Im}\left[\beta p^*\nabla p + \rho\mathbf{v}^*\cdot(\nabla)\mathbf{v}\right],$$

$$\mathbf{S} = \frac{\rho}{4\omega}\mathrm{Im}(\mathbf{v}^*\times\mathbf{v}), \quad \mathbf{L} = \mathbf{r}\times\mathbf{P}, \quad \mathbf{J} = \mathbf{L} + \mathbf{S}. \tag{2}$$

Here the spin density is defined as $\mathbf{S} = \dfrac{\rho}{4\omega}\text{Im}(\mathbf{v}^* \times \mathbf{v})$ to ensure the self-consistency of the theory and the conservation of angular momentum of the whole system, which are supported by the analytical derivation and experimental results below. The non-zero spin AM density is caused by the local rotation of the vector velocity field and can be more intuitively understood with the help of the polarization ellipse, which is traced by the temporal evolution of the real-valued velocity field at each point (see purple arrows in the waveguide in Fig.1a).

As another important kind of dynamic properties, the kinetic Abraham-type momentum and total AM densities are given by energy flux density

$$\boldsymbol{\mathcal{P}} = \dfrac{1}{2c^2}\text{Re}(p^*\mathbf{v}), \quad \boldsymbol{\mathcal{J}} = \mathbf{r} \times \boldsymbol{\mathcal{P}}. \tag{3}$$

The relation between the kinetic-Abraham and canonical-Minkowski momenta is $\boldsymbol{\mathcal{P}} = \mathbf{P} + \dfrac{1}{2}\nabla \times \mathbf{S}$, with the physical interpretation that the former describes the properties of acoustic pressure fields only, while the latter characterizes the properties of both the pressure and velocity modes.

## Acoustic cylindrical guided modes and their angular momenta

Different from acoustic vortex beams in free space, acoustic guided vortices hold unique properties in terms of AM due to the acoustic field redistribution executed by boundary. In a cylindrical waveguide of radius $R$ with pure reactance boundary, there exist undamped vortex modes, with the expression of acoustic pressure and velocity field in the cylindrical coordinates $(r, \varphi, z)$ as (the time factor $e^{-i\omega t}$ is omitted):

$$\begin{aligned}p(r,\varphi,z)\big|_{r \leq R} &= A J_l(k_r r)\exp[i(l\varphi + k_z z)],\\ \mathbf{v}(r,\varphi,z)\big|_{r \leq R} &= \dfrac{1}{i\rho\omega}\nabla p(r,\varphi,z),\end{aligned} \tag{4}$$

where $A$ is the amplitude, $l = 0, \pm 1, \pm 2, \ldots$ is the azimuthal quantum number, $J_l(r)$ is the $l$th-order Bessel function of the first kind, $k_r$ is a quantized radial wave number determined by boundary impendance $Z = \frac{p}{v_r}\Big|_{r=R}$, and $k_z = \sqrt{k^2 - k_r^2}$ is the propagation constant with $k = 2\pi/\lambda$ being the wavenumber.

The main dynamical properties of the cylindrical guided modes can be analytically derived. By substituting Eq. (4) into Eqs. (1) and (2) and integrating over any cross-sections perpendicular to the propagation axis, the canonical Minkowski-type spin, orbital, total AM and energy can be obtained as

$$\langle \mathbf{S} \rangle = \langle S_z \rangle \hat{\mathbf{z}} = \frac{\pi R^2 l}{2\omega} \beta |A|^2 \frac{1}{k^2 R^2} J_l^2(k_r R) \hat{\mathbf{z}},$$

$$\langle \mathbf{L} \rangle = \langle L_z \rangle \hat{\mathbf{z}} = \frac{\pi R^2 l}{2\omega} \beta |A|^2 \left[ \Psi - \frac{1}{k^2 R^2} J_l^2(k_r R) \right] \hat{\mathbf{z}},$$

$$\langle \mathbf{J} \rangle = \langle J_z \rangle \hat{\mathbf{z}} = \frac{\pi R^2 l}{2\omega} \beta |A|^2 \Psi \hat{\mathbf{z}},$$

$$\langle W \rangle = \frac{\pi R^2}{2} \beta |A|^2 \Psi, \qquad (5)$$

where $\langle ... \rangle = \iint_{r<R} ... dxdy$ is the integration in two dimensions (2D), and $\Psi = \left[1 - l^2/(k_r^2 R^2)\right] J_l^2(k_r R) + J_l'^2(k_r R) + k_r R/(k^2 R^2) J_l(k_r R) J_l'(k_r R)$. Notice that these AM are all along z-axis and have no component along x- or y-axes, so in the following derivation we focus only on the z-axis components. These components of AM can be normalized as

$$\bar{S}_z = \frac{\langle \psi | \hat{S}_z | \psi \rangle}{\langle \psi | \psi \rangle} = \frac{\omega \langle S_z \rangle}{\langle W \rangle} = \frac{l}{k^2 R^2 \Psi} J_l^2(k_r R),$$

$$\bar{L}_z = \frac{\langle \psi | \hat{L}_z | \psi \rangle}{\langle \psi | \psi \rangle} = \frac{\omega \langle L_z \rangle}{\langle W \rangle} = l \left[ 1 - \frac{1}{k^2 R^2 \Psi} J_l^2(k_r R) \right],$$

$$\bar{J}_z = \frac{\langle \psi | \hat{J}_z | \psi \rangle}{\langle \psi | \psi \rangle} = \frac{\omega \langle J_z \rangle}{\langle W \rangle} = l. \qquad (6)$$

Equation (6) is the first central result of this article, based on which we can analytically describe the normalized AM under various boundary conditions and deduce some important conclusions laying the cornerstone of AM conversion and SOI, as well be discussed in what follows. The normalized total AM is always an integer value and equal to the azimuthal quantum number *l*, while the magnitude of the normalized spin and orbital AM varies with the boundary conditions. The most important thing is that $\bar{S}_z$ is non-zero except for an absolutely-soft or infinite boundary (i.e., $J(k_r R) = 0$ or $R \to \infty$), suggesting that the guided acoustic vortices do carry spin AM aligning with propagating direction. Moreover, the normalized spin AM has continuous absolute value between 0 and 1 (i.e. $|\bar{S}_z| \in [0,1]$), resulting that it has the same direction with the orbital AM which is decided by the azimuthal quantum number $l$ and therefore $\text{sgn}(\bar{S}_z) = \text{sgn}(\bar{L}_z) = \text{sgn}(l)$.

Figures 1b and 1c demonstrate the detailed variation of spin and orbital AM when $l = +1$ according to Eq. (6), where abscissa $\theta = \sin^{-1}(k_r/k)$ characterizes the paraxial degree, and ordinate $kR$ characterizes the radial size of the guided modes. The yellow and white dashed lines are for the cases of the absolutely hard and the absolutely soft boundary, respectively, while the others are for the cases of pure reactance boundaries. It is not difficult to find in Figs. 1b and 1c that the positions of the absolutely soft boundary are where the minimum of the spin AM and the maximum of the orbital AM are located. And apart from the lowest one, the white lines slightly deviate from the position where the maximum of the spin AM and the minimum of the orbital AM are located due to the normalization. According to Eq. (2), the component of spin AM density along z-axis comes from the transverse velocity. Hence, here we demonstrate the polarization ellipse distribution of the velocity field $(v_x, v_y)$ of typical guided modes with different $l$ under three different boundary conditions in Fig. 2, where the greyscale background describes the acoustic pressure amplitude distribution and the ellipses characterize the projection of the velocity polarization on the cross-section, red for $S_z > 0$ and blue for $S_z < 0$ respectively. This offers qualitative physical images to interpret the Figs. 1b and 1c. When $l = +1$, the percentage of red ellipses is the largest when the boundary is absolutely hard, meaning that the part $S_z > 0$ dominates and the spin AM integrated over the whole cross-section is the largest. When the boundary is absolutely soft, the percentage of red and blue ellipses are comparable, and in this instance, the parts of $S_z > 0$ and $S_z < 0$ will offset each other after integrated over the whole cross-section, resulting in $\langle S_z \rangle = 0$, which is similar to the instance in free space; when the boundary is purely reactance, the percentage of red ellipses lies in between that under the above boundary conditions, and the red ellipses are slightly more than blue ones, and in this instance the absolute value of spatial integration of the part $S_z > 0$ is slightly more than that of the part $S_z < 0$, so $\langle S_z \rangle$ is also larger than 0 but smaller than that when boundary is absolutely hard. When $l = -1$, the situation of the percentage of blue ellipses is the same as red ones when $l = +1$.

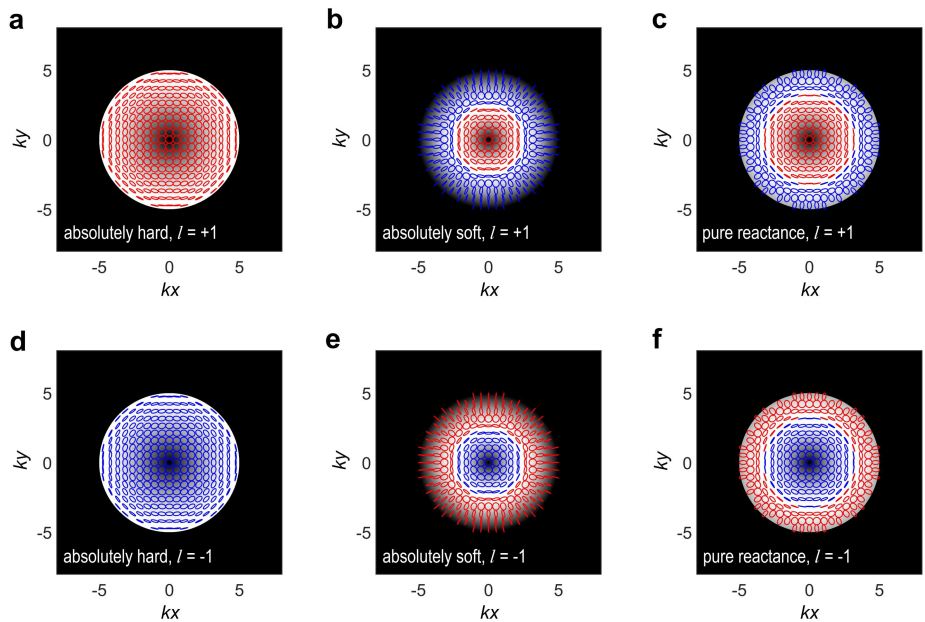

**Fig. 2| Distribution of the acoustic pressure amplitude (greyscale background) and polarization ellipses (red and blue ellipse) of the velocity field $(v_x, v_y)$ in the cross-sections with different $l$ under different boundary conditions. a**, absolutely hard, $l = +1$. **b**, absolutely soft, $l = +1$. **c**, absolutely hard, $l = +1$. **d**, absolutely hard, $l = -1$. **e**, absolutely soft, $l = -1$. **f**, absolutely hard, $l = -1$. The dimensionless radius $kR$

of the waveguide is 5, and red and blue correspond to right-handed ( $S_z > 0$ ) and left-handed ( $S_z < 0$ ) elliptical polarizations, respectively.

Next, we discuss the relationship between canonical-Minkowski and kinetic-Abraham quantities, which have long attracted considerable interest in optics yet remained unexplored in the context of acoustics. With $\mathbf{v}^* = \nabla p^* / (-i\rho\omega)$, the spin AM density in Eq. (2) can be further expressed as

$$\mathbf{S} = \frac{1}{2k^2} \nabla \times \mathcal{P}, \tag{7}$$

which reveals that the Minkowski-type spin AM density is associated with the curl of the Abraham-type momentum density. By further integrating both sides of Eq. (7) and considering the axial symmetry of the system, one has

$$\langle S_z \rangle = \frac{1}{2k^2} \oint_{r=R} \mathcal{P}_\varphi ds = \frac{\pi}{k^2} R \mathcal{P}_\varphi \bigg|_{r=R} \propto \mathcal{J}_z \bigg|_{r=R}. \tag{8}$$

Equation (8) is the second central result of this article, suggesting that the spin AM integrated over the whole cross-section is simply proportional to the kinetic-Abraham AM density at the edge of this cross-section. This substantially simplifies the spin AM observation procedure since one can readily calculate it from the acoustic field at the edge of the cross-section. Another point to note is the difference between the total AM values of these two types

$$\langle J_z \rangle - \langle \mathcal{J}_z \rangle = \int_{r<R} S_z dxdy - \frac{1}{2}\int_{r<R} \left[ \mathbf{r} \times (\nabla \times \mathbf{S}) \right]_z dxdy = \pi R^2 \, S_z \big|_{r=R}. \tag{9}$$

Equation (9) is the third central result of this article, for it reveals that the equivalence of canonical-Minkowski and kinetic-Abraham total AM depends on the spin AM density along the *z*-axis at the edge of the cross-section. When the boundary condition is absolutely hard or soft, one can derive $S_z \big|_{r=R} = 0$, hence, these two normalized total AM are equivalent (i.e. $\bar{J}_z = \bar{\mathcal{J}}_z$), the same as the case for free space. On the other hand, when the boundary condition is pure reactance, one has $S_z \big|_{r=R} \neq 0$, which means that these two variables are distinct, and their difference is proportional to the spin AM density at the edge (i.e., $\bar{J}_z - \bar{\mathcal{J}}_z \propto S_z|_{r=R}$). This also proves their difference in terms of total AM, namely: the normalized Minkowski-type total AM $\bar{J}_z$ is always an integer equal to the azimuthal quantum number $l$, while the normalized Abraham-type total AM $\bar{\mathcal{J}}_z$ may be different. Table 1 briefly summarizes the conclusions of AM properties and the relationship between canonical-Minkowski and kinetic-Abraham AM. Figure 3 shows the three kinds of normalized canonical-Minkowski AM and kinetic-Abraham total AM varying with the radius of cylindrical waveguide under different boundary conditions.

TABLE 1. Normalized AM Properties under Different Boundary Conditions

|  | $R \to \infty$ | Absolutely Hard | Absolutely Soft | Pure Reactance |
|---|---|---|---|---|
| Spin AM | $\bar{S}_z = 0$ | $\bar{S}_z \neq 0$, $\bar{S}_z \propto \mathcal{J}_z|_{r=R}$ | $\bar{S}_z = 0$ | $\bar{S}_z \neq 0$, $\bar{S}_z \propto \mathcal{J}_z|_{r=R}$ |
| canonical-Minkowski and kinetic-Abraham AM | $\bar{J}_z = \bar{\mathcal{J}}_z$ | $\bar{J}_z = \bar{\mathcal{J}}_z$ | $\bar{J}_z = \bar{\mathcal{J}}_z$ | $\bar{J}_z \neq \bar{\mathcal{J}}_z$, $\bar{J}_z - \bar{\mathcal{J}}_z \propto S_z|_{r=R}$ |

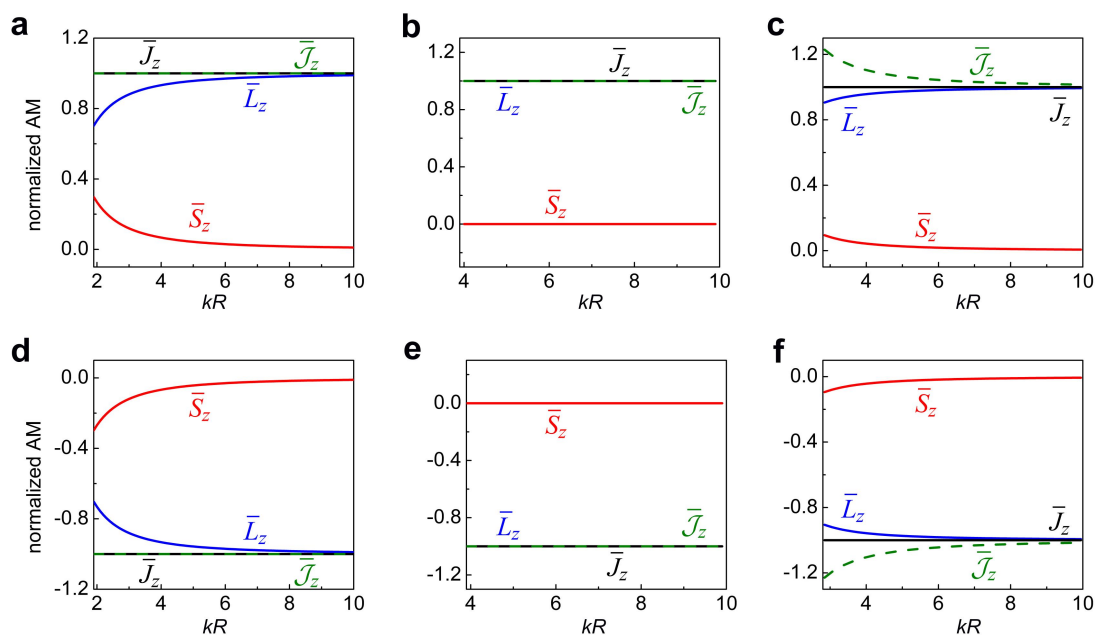

**Fig. 3| Numerical results of canonical-Minkowski spin, orbital, and total AM and the kinetic-Abraham total AM of the cylindrical guided modes with different $l$**

under different boundary conditions. **a**, $l = +1$, absolutely hard. **b**, $l = +1$, absolutely soft. **c**, $l = +1$, pure reactance. **d**, $l = -1$, absolutely hard. **e**, $l = -1$, absolutely soft. **f**, $l = -1$, pure reactance.

## Spin-orbit interaction

In free space, the integral spin AM of an acoustic vortex beam is constantly equal to 0, resulting that the SOI being unavailable for the missing spin degree of freedom. In cylindrical waveguides, in stark contrast, the analytical results we have demonstrated above validate that an acoustic vortex beam can carry non-zero integral spin AM when the boundary is absolutely hard or pure reactance. This means that SOI is possible to be realized as long as the boundary condition is reasonably designed even in a longitudinal wave system like acoustics. Here we design a waveguide with acoustically-hard boundary and slowly varying cross-section, which possesses the ability to support the spin and orbital AM conversion.

When $z = 0$, the radius of the cross-section is $R_0$, and the acoustic pressure of vortex beams propagating along $z$-axis in this cross-section can be expressed as $p(r,\varphi,0) = A_0 J_l\left(a_{l,m} r/R_0\right)\exp(il\varphi)$, where $a_{l,m}$ is the $m$th pole of $l$th-order Bessel function of the first kind and $A_0$ is a constant to characterize the amplitude. Let the radius of the arbitrary cross-section is $R(z)$, and $R(0) = R_0$. Then the pressure in the whole waveguide can be obtained

$$p(r,\varphi,z) = A(z) J_l\left[a_{l,m} r/R(z)\right]\exp\left\{il\varphi + iz\sqrt{k^2 - \left[a_{l,m}/R(z)\right]^2}\right\}, \tag{10}$$

where $|A(z)|^2 = |A_0|^2 R_0^2 \sqrt{k^2 - \left(a_{l,m}/R_0\right)^2} \Big/ \left\{R^2(z)\sqrt{k^2 - \left[a_{l,m}/R(z)\right]^2}\right\}$ results from the conservation of energy. Then the three kinds of normalized AM can be obtained by substituting Eq. (10) into Eq. (6) and applying Neumann boundary condition $\partial p/\partial r\big|_{r=R(z)} = 0$, as follows

$$\bar{S}_z(z) = \frac{\omega\langle S_z(z)\rangle}{\langle W(z)\rangle} = \frac{l}{k^2 R^2(z)\left(1-l^2/a_{l,m}^2\right)},$$

$$\bar{L}_z(z) = \frac{\omega\langle L_z(z)\rangle}{\langle W(z)\rangle} = l\left[1 - \frac{1}{k^2 R^2(z)\left(1-l^2/a_{l,m}^2\right)}\right],$$

$$\bar{J}_z(z) = \frac{\omega\langle J_z(z)\rangle}{\langle W(z)\rangle} = l. \qquad (11)$$

Equation (11) shows that the normalized spin and orbital AM vary with the coordinate along z-axis in this kind of waveguide, and more specifically, are negatively and positively correlated with the radius of the waveguide's cross-section, respectively; the total AM, however, does not change with coordinates and always equals to the azimuthal quantum number $l$. This means that this kind of waveguide prompts the mutual conversion of the spin and orbital AM during the propagating process, in other words, SOI occurs.

In experiment, we set $R(0)$ = 5cm, $R(z) = R(0) + 0.03z$, and the azimuthal quantum number $l$ = 1. The experimental system is shown in Fig. 4a. We measured the acoustic pressure distributions of four cylindrical areas around $z$ = 0.1, 0.3 and 0.7m, and then calculated the canonical-Minkowski spin, orbital and total AM density and their normalized results according to the measured acoustic pressure. The specific experimental details are provided in "Methods". The analytical (solid line), numerical (solid square) and experimental (solid circle) results of normalized AM are shown in Fig. 4b, from which we can clearly observe that theoretical prediction are adequately verified by both numerical and experimental results. As the radius of the waveguide increases with $z$, the normalized spin and orbital AM decreases and increases respectively with the normalized total AM remaining conserved, showing that a part of spin AM is transformed to orbital AM, which means that SOI is successfully observed in experiment. Figure 4c shows typical results of specific spin and orbital AM density distributions at some cross-sections. Insets 3 and 4 in Fig. 4c are slightly different from others at the central points. This is because the central point holds amplitude zero and phase singularity for a vortex beam, which greatly increases the difficulty of measurement and inevitably introduces errors, resulting in the discrepancy between experimental and analytical results. However, the error in the central point has negligible effect on the value of the normalized AM which is integrated over the whole cross-section.

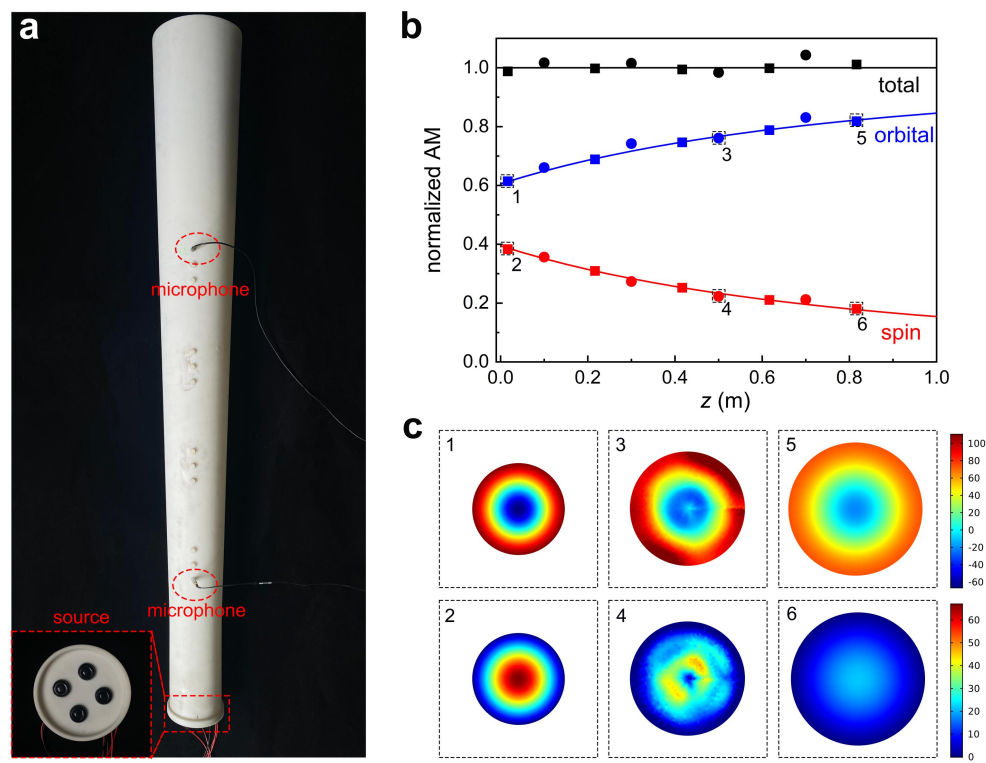

**Fig. 4| Experimental results of SOI observation. a**, Photograph of experimental system to observe SOI. **b**, Normalized canonical-Minkowski spin, orbital and total AM, and solid line, square and circle represents analytical, numerical and experimental results. **c**, Typical situations of spin and orbital AM density distribution at the cross-section selected from **b**, and the numbers 1 ~ 6 in two figures correspond to each other.

## Conclusion

To conclude, we have first analytically derived the acoustic canonical-Minkowski spin, orbital and total AM of the cylindrical guided modes by a self-consistent quantum-like representation and find that the spin AM is not equal to 0 when the boundary is absolutely hard or pure reactance, in stark contrast to the situation in the free space. Then we have analyzed the relationship between canonical-Minkowski and kinetic-Abraham total AM. We find that these two variables are not equivalent when the boundary is pure reactance, and the difference between them is proportional to the canonical-Minkowski spin AM density at the edge of the waveguide. More importantly, the normalized canonical-Minkowski total AM is conserved and strictly equal to the azimuthal quantum number *l*, while the kinetic-Abraham AM is not. At last, we have designed a waveguide with slowly varying cross-sections that can support SOI in longitudinal acoustic systems, which is previously considered unavailable. Experimental results verified that the SOI can occur in such a waveguide during propagating, with the evidence that a part of spin AM is transformed to orbital AM. Our research gives deep insight into the physics in acoustic and other longitudinal waves' AM study, and opens up possibility for phenomena previously thought to be impossible. This provides solid theoretical support for the development of spin and orbital AM-based acoustic devices, which may open up new degrees of freedom for underwater communications and object manipulations based on acoustic vortices.

## Methods

In experiment, the waveguide with a slow-varying cross-section, made of resin material with a wall thickness of 5mm, was fabricated using 3D printing technology. The vortex field was excited by four loudspeakers at the head with initial phases of 0, $0.5\pi$, $\pi$, and $1.5\pi$, respectively, at a frequency of 2070Hz. Sound-absorbing foam was placed at the end of waveguide to eliminate the undesired reflection. To investigate the velocity field in the waveguide, we conducted measurements of acoustic pressure along four truncated cylindrical regions in the *z*-direction. This allowed us to calculate the velocity field of the central cross-sections corresponding to these regions and subsequently calculate the three kinds of AM in these cross-sections. Considering that the large number of openings on the waveguide wall can affect the boundary conditions for sound propagation, we obtain the acoustic pressure at different azimuth angles by rotating the source array per measurement. The sensors we used in experiment were 1/4-inch free field microphones (BRÜEL & KJÆR Type 4961).

## Acknowledgement

This work was supported by the National Key R&D Program of China (Grant Nos. 2022YFA1404402 and 2017YFA0303700), the National Natural Science Foundation of China (Grant Nos. 11634006 and 12174190), High-Performance Computing Center of Collaborative Innovation Center of Advanced Microstructures and a project funded by the Priority Academic Program Development of Jiangsu Higher Education Institutions.

## Conflict of Interest

The authors declare no conflict of interest.